\documentclass[%
 reprint,
 aps,
 pre,
 amsmath,amssymb,
longbibliography
]{revtex4-2}

\usepackage[colorlinks,linkcolor=magenta,citecolor=magenta]{hyperref}

\usepackage{graphicx}
\usepackage{bm}
\usepackage{xcolor}

\usepackage{comment}
\begin{document}

\title{Experimental Modeling of Chiral Active Robots and a Minimal Model of Non-Gaussian Displacements}

\author{Yuxuan Zhou}
 \affiliation{
School of Physics and Astronomy, Yunnan University, South Section East Outer Ring Road, Chenggong District, Kunming, 650500, People's Republic of China}
 \author{Maomao Ge}
 \affiliation{
School of Physics and Astronomy, Yunnan University, South Section East Outer Ring Road, Chenggong District, Kunming, 650500, People's Republic of China}
\author{Ting Wang}%
\email{tingwang@ynu.edu.cn}
 \affiliation{
School of Physics and Astronomy, Yunnan University, South Section East Outer Ring Road, Chenggong District, Kunming, 650500, People's Republic of China}

\date{\today}

\begin{abstract}
We design 3D-printed motor-driven active particles and find that their dynamics can be characterized using the model of overdamped chiral active Brownian particles (ABPs), as demonstrated by measured angular statistics and translational mean squared displacements (MSDs). Furthermore, we propose a minimal model that reproduces the double-peak velocity distributions and further predicts a transition from the single-peak to the double-peak displacement distributions in short-time regimes. The model provides a clear physics picture of these phenomena, originating from the competition between the active motion and the translational diffusion. Our experiments confirm such picture. The minimal model enhances our understanding of activity-driven non-Gaussian phenomena. The designed particles could be further applied in the study of collective chiral motions.
\end{abstract}

\maketitle

\section{Introduction}
Active matter consists of self-propelled units and exhibits novel non-equilibrium dynamics~\cite{Marchetti2013,Bechinger2016rmp,chate2020dry}. Examples range from biological entities, such as cells~\cite{li2018intracellular,henkes2020dense}, bacteria~\cite{lauga2006swimming,zhang2010collective}, and animal collectives~\cite{ballerini2008interaction}, to human-engineered systems, such as active colloids~\cite{paxton2004catalytic,buttinoni2013dynamical}, shaker-driven active grains~\cite{kudrolli2008swarming,deseigne2010collective_disk_prl,deseigne2012vibrated_softmatter,Walsh2017noise_graular_softmatter,scholz2018inertial,sprenger2023inertial}, and self-propelled robots~\cite{deblais2018boundaries,dauchot2019_prl_bug_dynamics,baconnier2022_active_solid_natphys,siebers2023exploiting}. These systems exhibit rich non-equilibrium phenomena, including non-Gaussian diffusion~\cite{Zheng2013, scholz2018inertial,dulaney2020waves, Herrera2021_maxwellian_pre}, active pressure~\cite{solon2015pressure, Solon2015prl}, motility-induced phase separation~\cite{buttinoni2013dynamical, Cates2015mips_review, Shi2020prlbuble},
self-alignments of active solids ~\cite{baconnier2022_active_solid_natphys}, 
and as well as chirality induced phase separation~\cite{scholz2018} and odd viscosity~\cite{Dasbiswas2017oddviscosity,Souslov2019topological_wave,Liu2020,fruchart2023odd}.

These phenomena could emerge from microscopic models~\cite{solon2015active,martin2021statistical,shaebani2020computational,zottl2023modeling}. One of the most important models is that of active Brownian particles (ABPs), where a single particle moves along a given direction with constant speed, and the direction changes randomly due to rotational noise. The model is theoretically studied in~\cite{solon2015active,fily2012athermal,Solon2015prl,keta2024emerging}, experimentally realized in the microscopic systems of active colloidal particles\cite{Zheng2013,buttinoni2013dynamical,palacci2010sedimentation}, and in the macroscopic systems of shaker-driven active granular particles from overdamped dynamics~\cite{Walsh2017noise_graular_softmatter} to underdamped dynamics~\cite{scholz2018inertial,sprenger2023inertial}.

The experimental realizations above~\cite{Walsh2017noise_graular_softmatter,scholz2018inertial,sprenger2023inertial} are essentially achiral. For experimental studying of chiral ABPs (which, in addition to angular diffusion, exhibit fixed angular speeds), \textcolor{black}{the circular motion
of $L$-shaped micro-swimmers and their
interactions with boundaries
was studied in~\cite{kuemmel2013circular}},
the diffusion of shaker-driven chiral grass seeds in lattices was studied in~\cite{chan2024chiral}, where the diffusivity could be strongly affected by the lattices' geometry. The collective motion of commercially available light-driven walkers was recently studied in~\cite{siebers2023exploiting}. The mean squared displacements (MSDs) of a single walker indicate the validity of the overdamped chiral ABPs. But direct statistics of the angular dynamics
\textcolor{black}{for macroscopic systems} are absent, which are key characteristics of chirality. To what extend can the model of chiral ABPs be applied to other macroscopic systems is still unknown. In addition, neither grass seeds nor light-driven walkers are easy to modify, so robust and modifiable chiral ABPs are still needed in experiments.

Meanwhile, in contrast to Maxwellian velocity distributions of Brownian particles, the velocities distributions of ABPs may exhibit double-peak profiles as shown in~\cite{Zheng2013,scholz2018inertial}. Such non-Gaussian distributions were further studied in~\cite{Herrera2021_maxwellian_pre} based on the Fokker-Planck equation of underdamped chiral ABPs, in which inertia is crucial for the double-peak velocity distributions. However, this is probably not the whole story, as double-peak distributions have also been observed experimentally in active colloidal systems~\cite{Zheng2013} and theoretically calculated by expanding the Smoluchowski equation in orientational moments of the model of overdamped ABPs~\cite{dulaney2020waves}. In both cases, inertia is negligible. A clear physics picture of the double-peak velocities distributions is still missing. 

In this paper, we design 3D-printed electric motor driven particles, which encode both persistent and rotational motion. We are interested in two aspects: first\textcolor{black}{ly},  the robustness of the model of chiral ABPs, i.e. to what extent can the model describe our 3D-printed electric active particles, especially, in terms of angular dynamics statistics and the MSDs; second\textcolor{black}{ly}, can a minimal model be constructed to describe the double-peak velocity distributions and further predict short-time
displacement distributions? 

The paper is organized as follows: we introduce the experimental setup in Sec.~\ref{setup} and the model of overdamped chiral ABPs in Sec.~\ref{model}. Then we experimentally test the model by measuring the angular statistics in Sec.~\ref{angle_statics}, and the MSDs in Sec.~\ref{MSD}. Furthermore, we propose a minimal model to explain the double-peak velocity distributions found in our experiments in Sec.~\ref{minimal} and extend it to non-Gaussian displacements in Sec.~\ref{DD}. We conclude in Sec.~\ref{Conclude}.

\section{Experimental setup}\label{setup}
\begin{figure}[htbp]
    \includegraphics
    [width=0.425\textwidth]
        {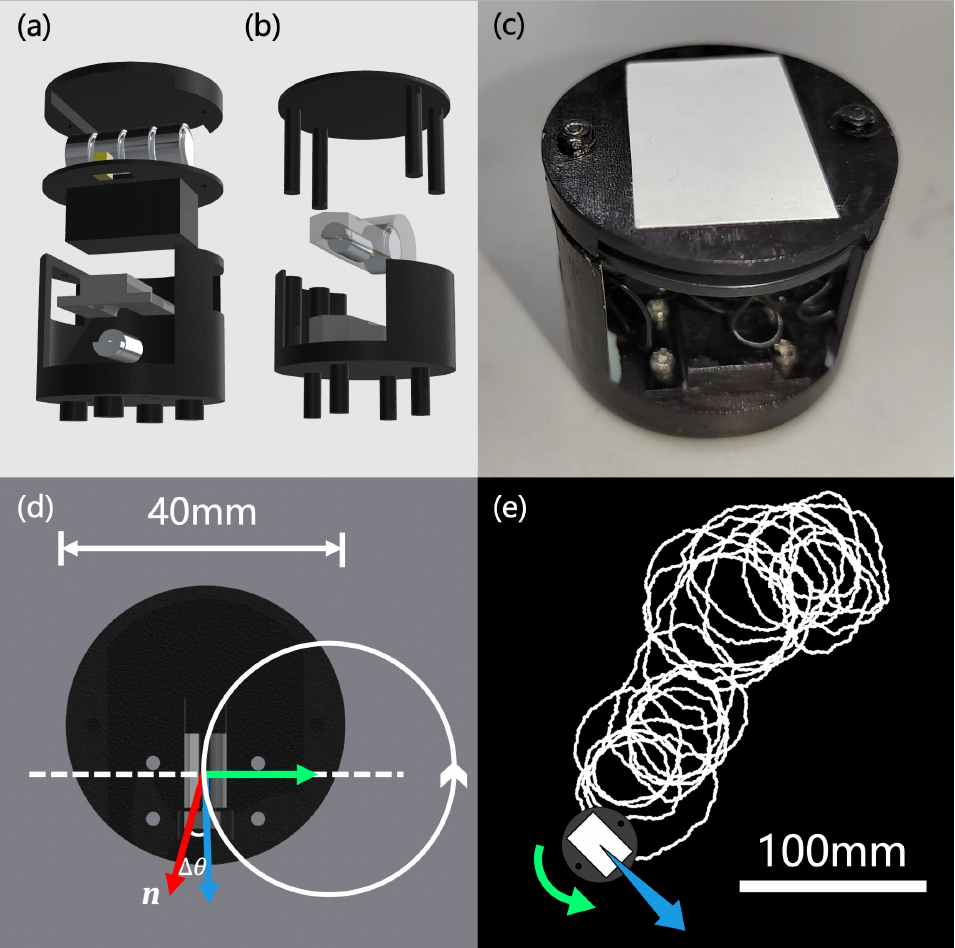}
    \caption{\label{fig:setup} Design of 3D-printed and motor driven active particles. 
    (a-b) Structures of a heavy and a light particle. (c) A physical picture of a heavy particle.  (d) Illustration of driven mechanism. (e) A trajectory recorded by a camera operating at 60fps.}
\end{figure}

We design 3D-printed, vibration motor driven active particles, see Fig.~\ref{fig:setup}. The particles are different in weights. The heavy particle has a weight  of 31.2g with four button batteries connected in parallel, see Fig.~\ref{fig:setup} (a). The light particle has a weight of 18.1g with one button battery only,  see Fig.~\ref{fig:setup} (b). 

The shells of the particles are 3D-printed with ABS thermosetting resin. Each particle is cylinder covered by a circular disc cover with a diameter of 40mm. At the bottom of each particle, four straight cylindrical legs are arranged in a square, two diagonal ones are on the same straight line with the long axis of a vibration motor inside, see Fig.~\ref{fig:setup} (a) and (b). A photo of the heavy particle is displayed in Fig.~\ref{fig:setup} (c). 

The detailed driven mechanism of our active particles is out of scope of this paper due to its complexity of non-linear dynamics~\cite{koumakis2016mechanism,scholz2016ratcheting}, which is essentially the centrifugal motion of the vibration motor causing two modes of motion, as shown in Fig.~\ref{fig:setup} (d). The vertical vibration with weight asymmetry enables the particles to behave as self-propelled polar "walkers"~\cite{kudrolli2008swarming,cheng2022experimental}, while the horizontal vibration induces a torque that makes the particles move in circles~\cite{vartholomeos2013analysis}. 

The positions and the angles of the active particles are captured by a camera operating at 60fps. A recorded trajectory of the heavy particle is shown in Fig.~\ref{fig:setup} (e).  

\section{Model}\label{model}
We explore the dynamics of our motor-driven chiral particles by the model of overdamped chiral ABPs as follows
\begin{subequations} \label{eom} 
\begin{eqnarray}
\bm{v}&=&v_0\bm{n}+\sqrt{2D_T}\bm{\xi}_r, \label{eom1}\\
\dot{\theta}&=& \Omega + \sqrt{2D_R} \xi_\theta. \label{eom2}
\end{eqnarray}
\end{subequations}
The total velocity $\bm{v}$ comprises both the self-propelled velocity and the effective thermal velocity induced by translational noise, where $v_0$ is the strength of the self-propelled velocity, $\bm{n}=(\cos\theta,\sin\theta)$ is its direction,  $D_T$ is the translational diffusivity, $\bm{\xi}_r=(\xi_{x},\xi_{y})$  is Gaussian white noise with zero mean and variance 
$\langle \xi_{\alpha}(t) \xi_{\beta}(t') \rangle=\delta_{\alpha,\beta}\delta (t-t'),\, \alpha,\beta\in \{x,y\}$. The total angular speed $\dot{\theta}$ comprises both the fixed angular speed $\Omega$
and the rotational noise, where $D_R$ is the rotational diffusivity, $\xi_\theta$ is Gaussian white noise with zero mean and variance $\langle \xi_\theta(t) \xi_\theta(t') \rangle=\delta (t-t')$. 
The dynamics are completely characterized by these four \textcolor{black}{physics} parameters: $v_0$, $D_T$, $\Omega$ and $D_R$ \textcolor{black}{\cite{ScalingComments}}.

\section{Angular statistics}\label{angle_statics}
\begin{figure*}
\includegraphics[width=0.9\linewidth]{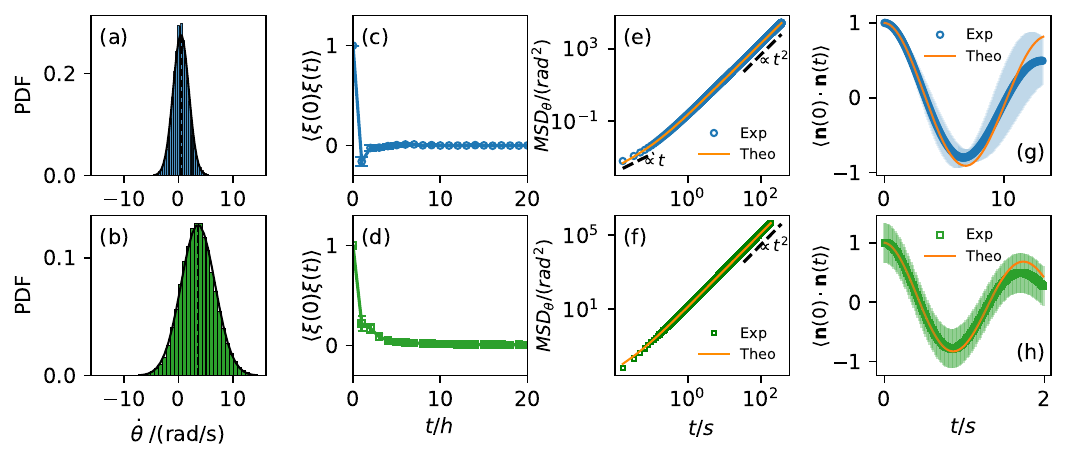}
\caption{\label{fig:angle}
Angular statistics of chiral particles.  
In the upper row are the results of the heavy particle, while in the lower, the results of the light one. (a) and (b) are the PDFs of the angular speed. Bars are experimental statistics. Solid black curves are normal distribution fitting. (c) and (d) are the noise ACFs. (e) and (f) are angular MSDs. Dotted points are experimental data. Solid orange lines are theoretical formula 
Eq.~\eqref{msd_angle}. (g) and (h) are 
ACFs of the directions. Dotted points 
with error bars are experimental data. Solid orange lines are theoretical formula Eq.~\eqref{director_acf}.}
\end{figure*}
The angular statistics of the two particles are plotted in Fig.~\ref{fig:angle}. In the upper row are the results of the heavy particle, while in the lower are the results of the light one. The probability distribution functions (PDFs) of their angular speed $\dot{\theta}$, are normally distributed, indicating that the noise is Gaussian, see Fig.~\ref{fig:angle} (a) and (b). The time auto correlation functions (ACFs) of the noises decay to zeros in a few time intervals of $h$ (
$h=1/60 s$ \textcolor{black}{due to the chosen 
highest recording frame rate of our camera, $60$Hz, 
} representing the time interval between \textcolor{black}{two} adjacent frames), see Fig.~\ref{fig:angle} (c) and (d). In order to extract the fixed angular speed $\Omega$ and the rotational diffusivity $D_R$, we fit the mean  $\langle\Delta \theta (t)\rangle=\Omega t$  and the variance $Var[\Delta \theta (t)]=2D_R t$ in short time, where $\Delta \theta (t):=\theta(t)-\theta(0)$ is the angular displacement. In practice, we fits the mean and the variance in the time interval $t=100 h$. The extracted $\Omega$ and $D_R$ for both heavy and light particles are recorded in Table~\ref{table1}. 

To further check the accuracy of the extracted $\Omega$ and $D_R$, we compute the MSD of the angle $\theta(t)$ and the ACF of the direction $\bm{n}(t)$ from the model of overdamped chiral ABPs, and compare them with the direct measurements.

Eq.~\eqref{eom2} indicates the MSD of the angle should be 
\begin{equation}\label{msd_angle}
MSD_\theta(t) = \Big\langle \Delta \theta(t)^2 \Big\rangle = \Omega^2 t^2 + 2D_R t,
\end{equation}
where we use the \textcolor{black}{decorrelation} of the noise $\langle \xi_\theta(t)\xi_\theta(t')\rangle=\delta(t-t')$. The MSDs of the angles can be well fitt  
   ed by the theoretical formula Eq.~\eqref{msd_angle}, where the dotted points are experimental data and the solid orange lines are the predictions from Eq.~\eqref{msd_angle} with 
measured $\Omega$ and $D_R$ in Table~\ref{table1}, 
see Fig.~\ref{fig:angle} (e) and (f). Note that in short time regimes, the angular MSD of the heavy particle is already in the linear regime, due to the domination of the angular diffusion term $2D_R t$. While in long time regimes, the angular MSDs of both particles are proportional to $t^2$, due to the domination of the fixed rotation term $\Omega^2 t^2$. 

According to Eq.~\eqref{eom2}, we compute the ACF of the direction $\bm{n}(t)$
\begin{equation} \label{director_acf}
   \Big\langle \bm{n}(t)\cdot\bm{n}(0) \Big\rangle =\Big\langle \cos\Delta \theta(t) \Big\rangle = e^{-D_R t}\cos\Omega t,
\end{equation}
where we have used the property that $\Delta \theta (t)$ is normally distributed with mean $\mu=\Omega t$ and variance $\sigma^2=2D_R t$. The ACFs of the directions for both heavy and light particles can be reasonably fitted by Eq.~\eqref{director_acf}. The small deviation may arise from fluctuations in $D_R$ and $\Omega$ due to the inevitable battery power instability.  
\section{Mean squared displacements}\label{MSD}
The complete model of chiral ABPs requires
the additional parameters: the self-propelled velocity $v_0$ and the translational diffusivity $D_T$, which are 
extracted from the PDFs of the particles' velocities being parallel to the angular direction, $f_{v_{\parallel}}(v)$. The PDFs for both two particles are found to be Gaussian, and the self-propelled velocities equal to the means, $v_0=\mu_{v_\parallel}$. The translational diffusivities relate to the variances as $D_T=2\sigma^2_{v_\parallel
}\textcolor{black}{/}h$. \textcolor{black}{The details of how to extract  $v_0$ and $D_T$ from the PDFs of $fv_{\parallel}$
is presented in Appendix~\ref{app}}. We collect all the parameters for both particles in Table~\ref{table1}.
\begin{table}[htbp]
    \centering    \caption{\label{table1}Parameters of chiral particles.}
    \begin{ruledtabular}
    \begin{tabular}{ccccc}
        & $\Omega \,[rad/s]$ & $D_R\,[rad^2/s]$ & $v_0\,[cm/s]$ & $D_T\,[cm^2/s]$\\
        \colrule
        heavy &  0.457$\pm$0.068& 0.0134 $\pm$0.0056& 1.04$\pm$0.10& 0.007$\pm$0.001\\ 
        light & 3.591$\pm$0.466& 0.220$\pm$0.092& 8.70$\pm$0.29& 0.037$\pm$0.009\\
    \end{tabular}
    \end{ruledtabular}
\end{table}

The translational MSD of a chiral ABP in ~Eq.~\eqref{eom} can be calculated analytically~\cite{van2008dynamics,weber2011correlated_noise,sevilla2020two}, 
\begin{equation}\label{msd}
\begin{aligned}
&\langle |\mathbf{r}(t) - \mathbf{r}(0)|^2 \rangle= \frac{2 \left(\frac{v_0}{D_R}\right)^2}{1 + \Gamma^2} \Big\{ -\frac{1 - \Gamma^2}{1 + \Gamma^2}+
D_R t+\\
& e^{-D_R t}\big[ \frac{1 - \Gamma^2}{1 + \Gamma^2} \cos(\Gamma D_R t) - \frac{2\Gamma}{1 + \Gamma^2} \sin(\Gamma D_R t) \big] \Big\}+4D_T t,\\
\end{aligned}
\end{equation}
where $\Gamma = \Omega/D_R$ represents the strength of the chirality 
($\Gamma=0$, achirality). In long time regimes,  the chiral ABP behaves diffusively with diffusivity 
\begin{equation}\label{D_eff}
D_{eff} = \lim_{{t \to \infty}} \frac{\langle|\mathbf{r}(t) - \mathbf{r}(0)|^2\rangle}{4t} = \frac{v_0^2}{2D_R(1 + \Gamma^2)}+D_T,
\end{equation}
while in short time regimes, the particle dynamics shifts from diffusion $D_Tt $ to ballistic motion $v_0^2 t^2$. 
\begin{figure}
    \centering
    \includegraphics[width=1\linewidth]{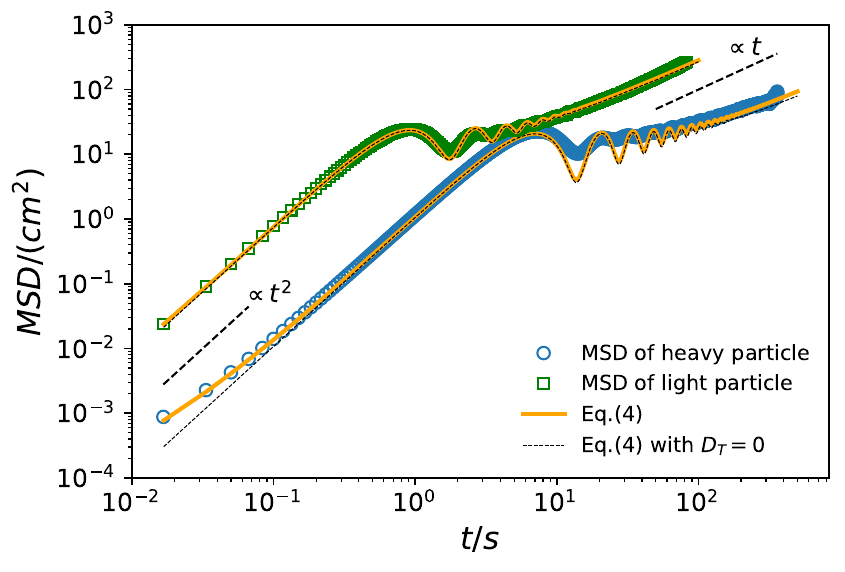}
    \caption{MSDs of heave and light particles. The solid orange lines 
    are the results from Eq.~\eqref{msd} with input values of $v_0,\Omega,D_R,D_T$
    in Table~\ref{table1}. The dashed black lines are the results from Eq.~\eqref{msd} with the same input values except for $D_T=0$.} \label{fig:msd}
\end{figure}

The experimental results of the translational MSDs of both heavy and light particles match the analytical expression Eq.~\eqref{msd} well, including the crossover from diffusion  $D_T t$ to ballistic motion $v_0^2 t^2$, see Fig.~\ref{fig:msd}, which supports the validity of the model of overdamped chiral ABPs for our experimental particles. Importantly, all the parameters input to Eq.~\eqref{msd} are directly from Table~\ref{table1} without any fitting. 

\textcolor{black}{
In addition, we can also extract $v_0$ and $D_T$ using the MSDs. 
However, the large statistical errors of the long-time MSDs prevent accurate extraction of $D_{eff}$ and consequently $D_T$. 
Fortunately, the short-time MSDs provide effective estimates. With minimal statistical errors in these regimes, we use experimental MSDs from $t=0$ to $t=20h$ to fit the MSD formula, Eq.~\eqref{D_eff}, treating $v_0$ and $D_T$ as least-squares fitting parameters, $D_R$ and $\Omega$ as input parameters obtained 
from each angle time series. For heavy and light particles, we obtained $v_0=1.07\pm 0.11 (cm/s)$,  
$D_T=0.006\pm0.001(cm^2/s)$
and $v_0=8.82\pm 0.30(cm/s)$, $D_T=0.034\pm 0.023(cm^2/s)$ respectively—values consistent with those extracted from the velocity distributions $fv_{\parallel}$.}

\textcolor{black}{
We further verify the output of velocity-PDFs in 
Fig. 6 and displacement-PDFs in Fig. 7 using the values of $v_0$ and $D_T$ obtained by this short-time MSDs method, and find no significant differences compared to the values extracted from the velocity distributions $fv_{\parallel}$. Thus, we will continue to use the parameters from Table~\ref{table1} in the following discussions. 
}

\section{Velocity distributions in laboratory coordinates}\label{minimal}
Let's consider the velocity distributions along an arbitrary direction in the laboratory coordinates. The velocity distributions are typically non-Gaussian. They may exhibit double-peak, which was found in \cite{Zheng2013,scholz2018inertial} and further treated as the inertial effect by Fokker-Planck equation of a single underdamped chiral ABP~\cite{Herrera2021_maxwellian_pre}. Note that the double-peak distributions found in \cite{Zheng2013} are of active colloids, whose dynamics are typically overdamped, 
suggesting that \textcolor{black}{inertia} is probably not the crucial point for the double-peak distributions. 

Based on the model of overdamped chiral ABPs, we construct a minimal model, which predicts a double-peak velocity distribution arising from the domination of the active velocity over the effective thermal velocity due to translational noise. The prediction is confirmed by the experimental results. 
\subsection{Minimal model}
The velocity along x-axis in Eq.~\eqref{eom1}   is the sum of  two independent components $v_x=v_A+v_T$; the active velocity $v_{A}=v_0\cos\theta$ and the effective thermal velocity $v_{T}=\sqrt{2D_T} \xi_x$. The PDF of the total velocity $f_{v_x}$ should be the convolution of the PDFs of the two independent components
\begin{equation}\label{fv_general}
    f_{v_x}(v)
= \int^{\infty}_{-\infty}
f_{v_A}(v')f_{v_T}(v-v') dv'.
\end{equation}
The PDF of the effective thermal velocity should be of normal distribution as
\begin{equation}\label{fvt}
    f_{v_T}(v)=\mathcal {N}(0,2D_T/h ),
\end{equation}
due to the white Gaussian noise
property $\xi_x=\frac{dW_t}{dt}\approx\frac{W_{t+h}-W_{t}}{h} \sim \mathcal{N}(0,1/h) $.  \textcolor{black}{
We denote $X\sim \mathcal{N}(\mu,\sigma^2)$,
if $X$ is normally distributed with 
mean $\mu$ and variance $\sigma^2$.
}
$h$ is the time resolution in experiment.  $W_t$ is a standard Wiener process \cite{calin2015informal} with vanished initial value $W_0=0$ and independent and normal distributed increment $W_{t_1}-W_{t_0}\sim \mathcal{N}(0,t_1-t_0)$ for any $t_1>t_0\geq0$.  

Let's consider the PDF of the active velocity $f_{v_A}$.  Due to the angular diffusion $\sqrt{2D_R}\xi_\theta$ and fixed orientation speed $\Omega$, the angle should be uniformly distributed, $\theta \sim$ uniform $(0,2\pi)$. The corresponding PDF of $\cos\theta$ is 
$f_{\cos\theta}(x)=\frac{1}{\pi \sqrt{1-x^2}}, for \,x \in [-1,1]; = 0,\, else.$  Thus the PDF of the active velocity $v_A=v_0\cos\theta$  should be 
\begin{equation}\label{fva}
f_{v_A}(v)= \begin{cases}
\frac{1}{v_0\pi \sqrt{1-(v/v_0)^2}},\quad &v \in [-v_0,v_0], \\
0,\quad & else.
\end{cases}
\end{equation}
Inserting Eq.~\eqref{fvt} and Eq.~\eqref{fva} into Eq.~\eqref{fv_general}, we get 
\begin{equation}\label{fv}
   f_{v_x}(v)= 
   \frac{1}{v_0}\int^{1}_{-1}
\frac{1}{\pi \sqrt{1-u^2}}
\frac{1}{\sqrt{2\pi} \sigma_v}
e^{-\frac{(\frac{v}{v_0}-u)^2}{2\sigma_v ^2}} du,
\end{equation}
where  we have denoted $v_{th}=\sqrt{\frac{2D_T}{h}}$  as the strength of the effective thermal velocity and 
\begin{equation}\label{sigma_v}
    \sigma_v=\frac{v_{th}}{v_0}=\frac{\sqrt{2D_T/h}}{v_0}
\end{equation}
as the ratio of the effective thermal velocity to the active velocity. 

One can readily compare the numerical integrals of Eq.~\eqref{fv}  with the experimental results.  Before that, let's discuss two limiting regimes to understand the distribution profiles. 
\begin{itemize}
    \item For $\sigma_v\gg 1$, the effective thermal velocity dominates over the active velocity. 
\begin{equation}
\begin{split}
  f_{v_x}(v)
&=\frac{1}{v_0}\int^{1}_
{-1}
\frac{1}{\pi \sqrt{1-u^2}}
\frac{1}{\sqrt{2\pi} \sigma_v}
e^{-\frac{(\frac{v}{v_0\sigma_v}-\frac{u}{\sigma_v})^2}{2}} du
\\
&\approx\frac{1}{v_0}\int^{1}_
{-1}
\frac{1}{\pi \sqrt{1-u^2}}
\frac{1}{\sqrt{2\pi} \sigma_v}
e^{-\frac{v^2}{
2(v_0\sigma_v)^2
}} du
\\
&=
\frac{1}{\sqrt{2\pi} (v_0\sigma_v)}e^{-\frac{v^2}{
2(v_0\sigma_v)^2
}} 
\int^{1}_
{-1}
\frac{1}{\pi \sqrt{1-u^2}}
du
\\
&=
f_{v_T}(v)
\end{split}
\end{equation}
where we have dropped the term $u/\sigma_v$ in $e^{-(\frac{v}{v_0\sigma_v}-\frac{u}{\sigma_v})^2/2}$ due to the fact
that $u/\sigma_v\ll 1$ in the case of $u\in [-1,1]$ and $\sigma_v\gg 1$.
Thus, as the effective thermal velocity dominates, the distribution of the velocity should be close to the PDF of the  effective thermal velocity.
    \item For $\sigma_v\ll 1$, the active velocity dominates over the effective thermal velocity. We have $\frac{1}{\sqrt{2\pi} \sigma_v}
e^{-\frac{(\frac{v}{v_0}-u)^2}{2\sigma_v ^2}}
\rightarrow \delta (u-\frac{v}{v_0})$ and 
\begin{equation}
  f_{v_x}(v)
\rightarrow\frac{1}{v_0}\int^{1}_{-1}
\frac{1}{\pi \sqrt{1-u^2}}
\delta (u-\frac{v}{v_0}) du
=f_{v_A}(v).
\end{equation}
Thus, as the active velocity dominates, the distribution of the velocity should be close to the PDF of the active velocity. 
\end{itemize}
\begin{figure}[htbp]
    \centering
    \includegraphics[width=1\linewidth]{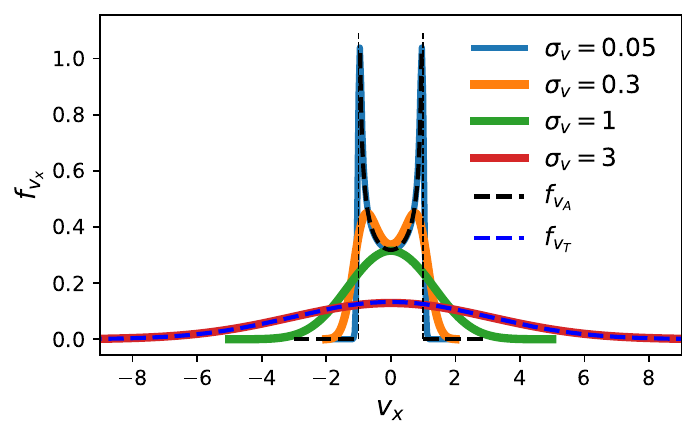}
    \caption{\label{fig_fv}Velocity distribution along x-axis. The solid lines are numerical integrals of Eq.~\eqref{fv} with different $\sigma_v=v_{th}/v_0$ (here we set $v_0=1$ without loss of generality). Dashed lines are the PDFs
    of $f_{v_A}$ and $f_{v_T}$ indicated by black and blue, respectively.}
\end{figure}

From the discussions above, 
we can deduce that the single-peak/double-peak distributions are determined by the relative strength of effective thermal velocity to the active velocity, $\sigma_v$ in Eq.~\eqref{sigma_v}. This finding is consistent with the numerical integration of Eq.~\eqref{fv} with varying values of $\sigma_v$, see Fig.~\ref{fig_fv}.

\subsection{Experimental results}
\begin{figure}[hptb]
\includegraphics[width=0.9\linewidth]{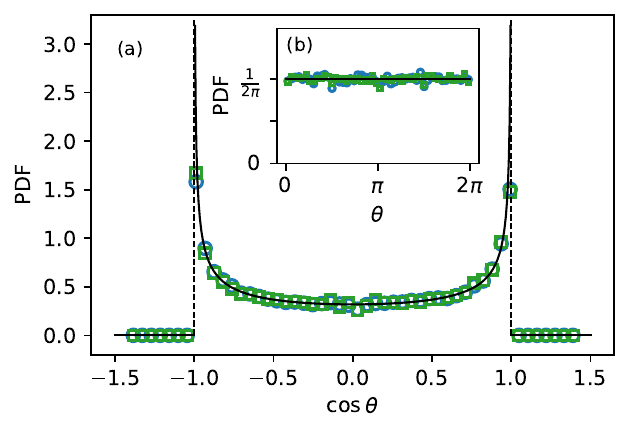}
        \caption{\label{fig:pdf_theta} PDFs of of angles. 
        Dots are experimental data of the PDF of $\cos\theta$ in (a) and the PDF of $\theta$ in (b). The solids lines 
        are the theoretical predictions: for the PDF of $\cos\theta$, $f_{\cos\theta}(x)=\frac{1}{\pi \sqrt{1-x^2}}, for \,x \in [-1,1]; = 0,\, else.$ For the PDF of $\theta$, 
        $f_\theta(x)=\frac{1}{2\pi}$,uniformly distributed.}
\end{figure}

Let's first check the angular distributions. We find that the angular distributions for both heavy and light particles are indeed uniform, see Fig.~\ref{fig:pdf_theta} (b). 
The corresponding PDFs of $\cos\theta$ also fit the theoretical prediction, see Fig.~\ref{fig:pdf_theta} (a). 

\begin{figure}[htbp]
    \centering
    \includegraphics[width=0.95\linewidth]{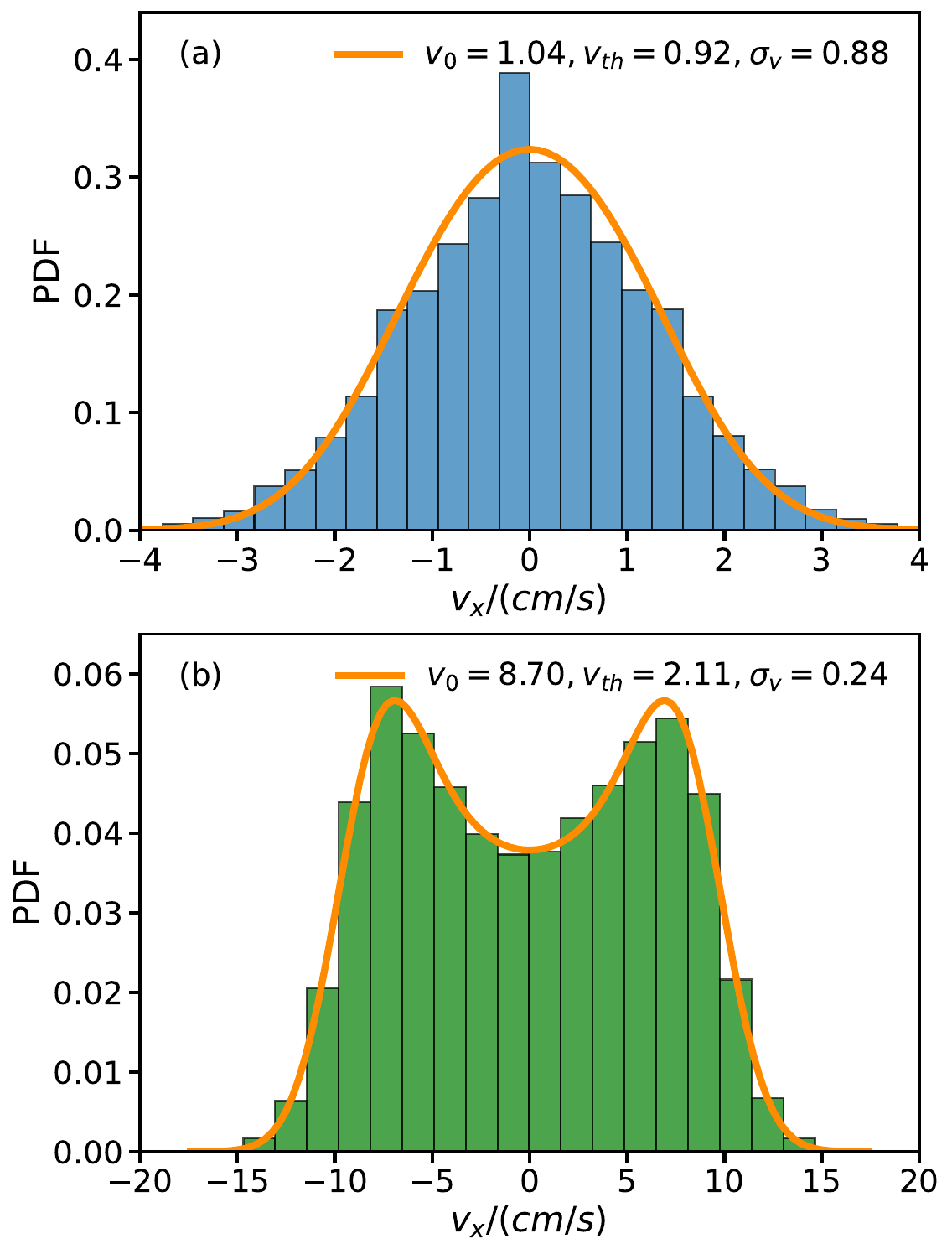}
        \caption{ \label{fig:pdf_v_exp}Experimental data of velocity distribution along x-axis. The PDF of the velocity of heavy
        particle exhibits single peak, as shown in histogram in (a).
        The PDF of the velocity of light
        particle exhibits double peak, as shown in histogram in (b).
        The orange solid lines are the theoretical predictions by Eq.~\eqref{fv}, fitting the experimental PDFs, respectively.  $v_0$ and $v_{th}
        =\sqrt{2D_T/h}$  are extracted from Table~\ref{table1}. $\sigma_v=v_0/v_{th}$ indicates the relative strenght of the active velocity over the translational noise velocity.}
\end{figure}

Then let's test the velocity distributions. As shown in Fig.~\ref{fig:pdf_v_exp}, the velocity distributions of the heavy particle exhibit single-peak, while the distributions of the light particle exhibit double-peak. Both of them can be well fitted by the minimal model, Eq.~\eqref{fv}. The parameters of $v_0$ and $v_{th}$ are directly extracted from Table~\ref{table1} without any fitting.

\section{DISPLACEMENT DISTRIBUTIONS}\label{DD}
        \begin{figure*}[htbp]
\includegraphics[width=1\linewidth]{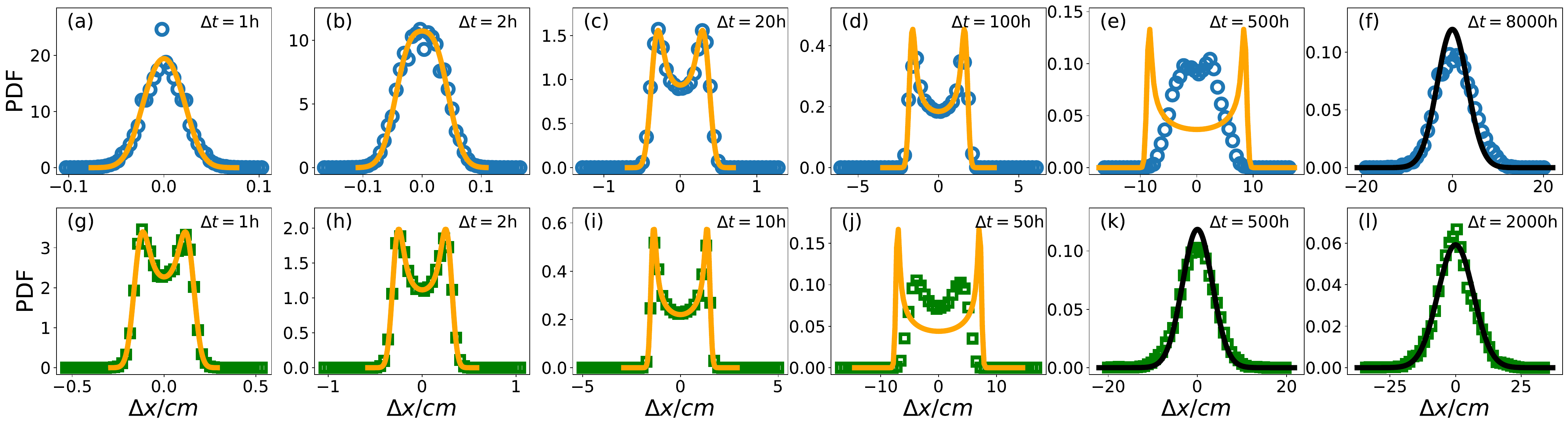}
    \caption{ \label{fig:pdf_dx}PDFs of the displacements for heavy (upper) and light (lower) particles. Dots are experiment data. Solid orange lines are  the results from Eq.~\eqref{f_dx} with input values of time interval $\Delta t$ and $v_0$,$D_T$ in Tabel~\ref{table1}. Solid black lines are the results of  
    the PDFs of long time diffusion 
    $e^{-\Delta x^2/(4 D_{eff}\Delta t)}/\sqrt{4\pi D_{eff}\Delta t}
    $
    with 
    the diffusivity in Eq.~\eqref{D_eff}.
   }
\end{figure*}
The minimal model can be directly extended to the displacement distributions in short time regimes, since the short-time displacement is 
\begin{equation}\label{dx}
    dx=v_xdt=v_0\cos\theta dt+\sqrt{2D_T}\xi_x dt.  
\end{equation}
The PDF of the displacements
should be 
\begin{equation}\label{f_dx}
   f_{dx}(\Delta x)
   =\frac{1}{v_0 \color{black}{\Delta t} }\int^{1}_{-1}
\frac{1}{\pi \sqrt{1-u^2}}
\frac{1}{\sqrt{2\pi} \sigma_{dx}}
e^{-\frac{(\frac{\Delta x}{v_0 \color{black}{\Delta t} }-u)^2}{2\sigma_{dx} ^2}} du,
\end{equation}
where the ratio of the effective thermal displacement to the
active displacement is
\begin{equation}\label{sigma_dx}
\sigma_{dx}=\frac{\sqrt{2D_T \color{black}{\Delta t} }}{v_0 \color{black}{\Delta t} }
    =\frac{\sqrt{2D_T /\color{black}{\Delta t} }}{v_0}.
\end{equation}
Interestingly,  Eq.~\eqref{sigma_dx} indicates that the PDF of the displacements could exhibit a crossover from the single-peak to the double-peak as the time interval increases. 
This could occur because the strength of the diffusion motion due to translational noise increases as $\propto \sqrt{\color{black}{\Delta t} }$, while the strength of the active motion increases as $\propto \color{black}{\Delta t} $. Consequently, the increase in time intervals effectively weakens the translational noise. Indeed, the crossover is observed in our experiments. The PDF of the displacements for the heavy particle clearly show the  crossover from the single-peak to the double-peak until the time interval $\Delta t=100h$
, see Fig.~\ref{fig:pdf_dx} (a)-(d). The dots are experimental results, matching well with the numerical integration results from Eq.~\eqref{f_dx}, as indicated in solid orange lines. 

For the long time intervals, Eq.~\eqref{f_dx}  does not match the experimental data, because the displacement formula Eq.~\eqref{dx} is valid only for the short time intervals\textcolor{black}{~\cite{ballistic_diffu1}}. For the long time intervals, the PDFs of the displacements should become purely diffusive, which 
is also confirmed in our experiments. The PDFs of the displacements 
in Fig.~\ref{fig:pdf_dx} (f) and (I) are fitted by the Gaussian distribution $\mathcal{N}(0, 2D_{eff} \Delta t)$, as indicated by the solid black lines, where $D_{eff}$ is the long time diffusivity in Eq.~\eqref{D_eff}\textcolor{black}{\cite{intermediate}}.

\section{conclusions and discussion}\label{Conclude}
We design 3D-printed motor driven active granular particles and find the particle dynamics can be characterized by the model of overdamped chiral ABPs, for both the rotational and translational MSDs. Then, we construct a minimal model that reproduces the double-peak velocity and displacement distributions. Our experiments validate the model. 

A few remarks should be addressed. \textcolor{black}{Firstly}, the minimal model is simple but quite useful. It not only 
reproduces the non-Gaussian 
diffusion phenomena, 
but also provides a clear 
mechanism: the interplay between the translational noise and active velocity leads to the observed crossover of single/double-peak distributions.

Second\textcolor{black}{ly}, 
\textcolor{black}{although chirality trajectories and 
the self-propelled directions
are discussed in the micro-swimmers~\cite{kuemmel2013circular},}
to the best of our knowledge, the \textcolor{black}{following} key angular statistical features of chirality\textcolor{black}{:} the quadratic term $\Omega^2 t^2$ in the long time angular MSDs, see Fig.~\ref{fig:angle} (e)
and (f), and the oscillation term $\cos\Omega t$ in the ACFs of the directions, 
see Fig.~\ref{fig:angle} (g)
and (h), have not been reported in experiments.

Third\textcolor{black}{ly}, the translational noise, often overlooked, is important as observed in our experiments. It not only causes a deviation from the expected $v_0^2t^2$ scaling in the MSD of the heavy particle in short time regimes, see Fig.~\ref{fig:msd}, but also plays a critical role in the formation of the non-Gaussian distributions.

To summarize, our experiment using self-desiged, 3D-printed,
motor-driven particles verifies the model of overdamped chiral ABPs.  Our minimal model accurately describes the experimental results, and offers a clear physical picture of active induced non-Gaussian diffusion. Using our particles, future studies could explore how particle interactions and inertia modify this picture.

\begin{acknowledgments}
We thank valuable discussions 
with Guangpeng Zhang, Ning Zheng and Thomas Speck. We acknowledge funding from the National Natural Science Foundation of China (Grants No. 12364031 and No. 11847070) and the Opening Project of Shanghai Key Laboratory of Special Artificial Microstructure Materials and Technology (Grant No. Ammt2022B-4). Yuxuan Zhou acknowledges support from the National Undergraduate Training Program for Innovation and Entrepreneurship \& Student Research Training Program. 
\end{acknowledgments}
\appendix*
\section{
    \textcolor{black}{
Extraction of $v_0$ and $D_T$ 
from Velocity Distributions along Self-Propelled Directions
    }
}\label{app}
\textcolor{black}{
We extract $v_0$ and $D_T$ in Table~\ref{table1} from the PDFs of velocities parallel to the self-propelled directions. According to Eq.~\eqref{eom1}, the velocities parallel and perpendicular to the self-propelled directions are given by:
}
\textcolor{black}{
\begin{subequations} \label{vp} 
\begin{eqnarray}
v_\parallel&=&\bm{v}\cdot \bm{n}=v_0+\sqrt{2D_T}\xi_{r,_\parallel}, \label{vp1}\\
v_\perp&=&\bm{v}\cdot \bm{n}_\perp=\sqrt{2D_T}\xi_{r,_\perp}\label{vp2}
\end{eqnarray}
\end{subequations}
where $\bm{n}_\perp=(\cos(\theta+\pi/2),\sin(\theta+\pi/2))$
is the direction perpendicular to the self-propelled direction $\bm{n}=(\cos\theta,\sin(\theta))$.  $\xi_{r,_\parallel}=\bm{\xi}_r\cdot \bm{n} $
and $\xi_{r,_\perp}=\bm{\xi}_r\cdot \bm{n}_\perp $
are the parallel and perpendicular components of 
the translational noise, which are Gaussian white noises,
leading to normal distributions of the velocities:
$v_\parallel\sim \mathcal{N}(\mu_{v_\parallel} = v_0,
\sigma^2_{v_\parallel} =2D_T/h)$ and 
$v_\perp\sim \mathcal{N}(\mu_{v_\perp} = 0,
\sigma^2_{v_\perp} =2D_T/h)$.
}
\begin{figure}[htbp]
\includegraphics[width=0.46\textwidth]{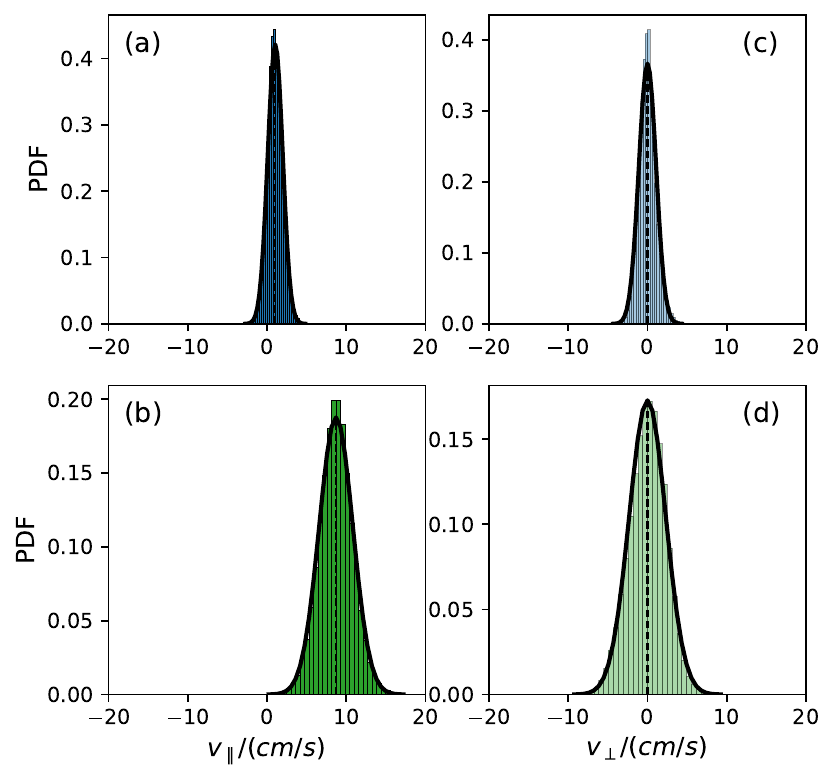}
     \caption{
\label{fig:fv12}
\textcolor{black}{
PDFs of the velocities
     for heavy (upper) and light (lower) particles. 
     (a) and (b) are PDFs 
     of the velocities being parallel 
     to the active force directions, while 
     (c) and (d) are the velocities being perpendicular
     to these directions. 
     Histograms are experiment data. 
     Solid black lines are  the 
     normal distributions fitting.}}
 \end{figure}
 
\textcolor{black}{Experimentally, we calculate velocities from the displacements between adjacent frames through $\bm{v}(t)=(\bm{r}(t+h)-\bm{r}(t))/h$, then project them along and perpendicular to the self-propelled directions to obtain $v_\parallel$ and $v_\perp$. In practice, we find the self-propelled direction is slightly misaligned with the motor's long axis, as shown by the blue and red arrows in Fig.~\ref{fig:setup}(d).  
Using the property that the average velocities
along the self-propelled direction should be $v_0$, 
and the perpendicular average velocities should vanish, we correct the angle using $\Delta \theta=\arctan{(\langle v_\perp\rangle/\langle v_\parallel\rangle)}$,
where $\langle \cdots\rangle$ indicates the statistical average
over all the trajectories. For heavy and light particles, $\Delta \theta$ is found to be $-23^\circ$ and $-14.1^\circ$, respectively. The corrected direction is then $\bm{n'} = (\cos \theta', \sin \theta')$ with $\theta' = \theta + \Delta \theta$, see the red arrow in Fig.~\ref{fig:setup}(d).}

\textcolor{black}{As shown in Fig.~\ref{fig:fv12}, the velocity histograms fit normal distributions: $f_{v_\parallel}^{fit}=\mathcal{N}(\hat{\mu}_{v_\parallel},\hat{\sigma}_{v_\parallel}^2)$ and $f_{v_\perp}^{fit}=\mathcal{N}(\hat{\mu}_{v_\perp},\hat{\sigma}_{v_\perp}^2)$. For each trajectory, $v_{\parallel}$ and $v_{\perp}$ are normally distributed, from which we extract $v_0=\hat{\mu}_{v_\parallel}$ and $D_T=\hat{\sigma}_{v_\parallel}^2/h$. The ensemble averages and standard deviations of $v_0$ and $D_T$ are shown in Table~\ref{table1} with statistical errors.
}

\bibliography{ref.bib}

\begin{thebibliography}{52}
\bibitem [{Sca()}]{ScalingComments}%
  \BibitemOpen
  \href@noop {} {}\bibinfo {note} {\textcolor{black}{After rescaling time and length by the persistent time $t_0=1/D_R$ and the persistent length $L_0=v_0/D_R$, the dimensionless form of Eq.~\eqref{eom} becomes $\bm{v'}(t')=\bm{n}(\theta' (t'))+\frac{\sqrt{2}}{Pe} \bm{\xi}'_r(t')$ and $d\theta'(t')/dt'=\Gamma + \sqrt{2}\xi'_\theta (t')$, where the relevant dimensionless parameters are the chirality strength $\Gamma=\Omega/D_R$ and the Peclet number $Pe=\frac{v_0}{\sqrt{D_T D_R}}$. The parameters $D_R$ and $v_0$ are scaled out due to the time rescaling $1/D_R$ and length rescaling $v_0/D_R$. We keep the dimensional form of Eq.~\eqref{eom} to effectively convey the scale of the physical parameters.}}\BibitemShut {Stop}%
\bibitem [{bal()}]{ballistic_diffu1}%
  \BibitemOpen
  \href@noop {} {}\bibinfo {note} {\textcolor{black}{ The choice of equations is based on the different dynamical behaviors of the system, which can be distinguished by the dimensionless time scale ratio $t' = t / t_0$, where $t_0$ is the persistent time defined in~\cite{ScalingComments}. For short times, $t'=tD_R\ll 1$, the system exhibits ballistic behavior: the displacements follow $dx=v_xdt$ (Eq. (13) ), and the PDFs of the displacements are govern by Eq. (14). For long times, $t'=tD_R\geq 1$, the dynamics shift to diffusive regimes, as the factor $e^{-D_Rt}$ in Eq. (4) suppresses transient dynamics, making Eq. (5) applicable.}}\BibitemShut {Stop}%
\bibitem [{int()}]{intermediate}%
  \BibitemOpen
  \href@noop {} {}\bibinfo {note} {\textcolor{black}{ For intermediate times, such as at $t = 500h$ (see Fig.~\ref{fig:pdf_dx} (f) and (i)), we find that $t' = tD_R^{\text{heavy}} \approx 0.1$ for heavy particles and $t' = tD_R^{\text{light}} \approx 2.0$ for light particles. This indicates that heavy particles remain in the ballistic regime, while light particles have transitioned to the diffusive regime. Consequently, we apply Eq. (14) for heavy particles and Eq. (5) for light particles at this time point. Fairly speaking, Eq. (14) does not fully capture the experimental distributions shown in Fig.~\ref{fig:pdf_dx} (e). To satisfy the ballistic regime, both $tD_R \ll 1$ and $t\Omega \ll 1$ must hold. The latter condition implies that the time must be short enough for the fixed rotation to have minimal impact on the particle’s persistent motion.}}\BibitemShut {Stop}%
\end{thebibliography}

\end{document}